\documentclass[aps,prl,twocolumn,showpacs,floatfix]{revtex4}
\usepackage{amsmath,bm,epsfig}
\usepackage{epsfig,color}
\usepackage{bm} 
\usepackage{ulem}

\newcommand{\C}[1]{{\mathcal{#1}}}
\renewcommand{\it}[1]{\textit{#1}}

\begin{document}

\title{Population dynamics at high Reynolds number}
\author{Prasad Perlekar$^{(1)}$, Roberto Benzi$^{(2)}$, David
  R. Nelson$^{(3)}$, Federico Toschi$^{(1)}$}

\affiliation{$^{(1)}$ Department of Physics, and Department of
  Mathematics and Computer Science, and J.M. Burgerscentrum, Eindhoven
  University of Technology, 5600 MB Eindhoven, The Netherlands; \\and
  International Collaboration for Turbulence Research. \\ 
  $^{(2)}$ Dip. di Fisica and INFN, Universit\`a ``Tor
  Vergata", Via della Ricerca Scientifica 1, I-00133 Roma, Italy. \\
  $^{(3)}$ Lyman Laboratory of Physics, Harvard University,
  Cambridge, MA 02138, USA}

\begin{abstract}
  We study the statistical properties of population dynamics evolving
  in a realistic two-dimensional compressible turbulent velocity
  field.  We show that the interplay between turbulent dynamics and
  population growth and saturation leads to quasi-localization and a
  remarkable reduction in the carrying capacity. The statistical
  properties of the population density are investigated and quantified
  via multifractal scaling analysis.  We also investigate numerically
  the singular limit of negligibly small growth rates and
  delocalization of population ridges triggered by uniform advection.
\end{abstract}

\pacs{47.27.-i, 47.27.E-, 87.23.Cc \hspace{0.5cm} Keywords: Fisher
  equation, population dynamics, turbulence}

\maketitle

For high nutrient concentration on a hard agar plate, the Fisher
equation \cite{fisher} can be a good description of the spreading of
microorganisms such as bacteria at low Reynolds number
\cite{Wakika}. However, many microorganisms, such as those in the
ocean, must find ways to thrive and prosper in high Reynolds number
fluid environments. In presence of a turbulent advecting velocity
field ${\bf u}({\bf x},t)$, the Fisher equation reads
\begin{equation} 
  {\frac{\partial C}{\partial t}} + {\bm \nabla}
  \cdot ({\bf u}C) = D \nabla^2 C + \mu C(1 - C),
\label{eq:fish}
\end{equation}
where $C({\bf x},t)$ is a continuous variable describing the
concentration of micro-organisms, $D$ is the diffusion coefficient and
$\mu$ is the growth rate. As an example of "life at high Reynolds
number", we could take Eq.~(\ref{eq:fish}) to represent the density of
the marine cyanobacterium Synechococcus \cite{moo95} under conditions
of abundant nutrients, so that $\mu \sim $ constant.

\begin{figure*}[!t]
  \begin{center}
    \includegraphics[width=0.45\linewidth]{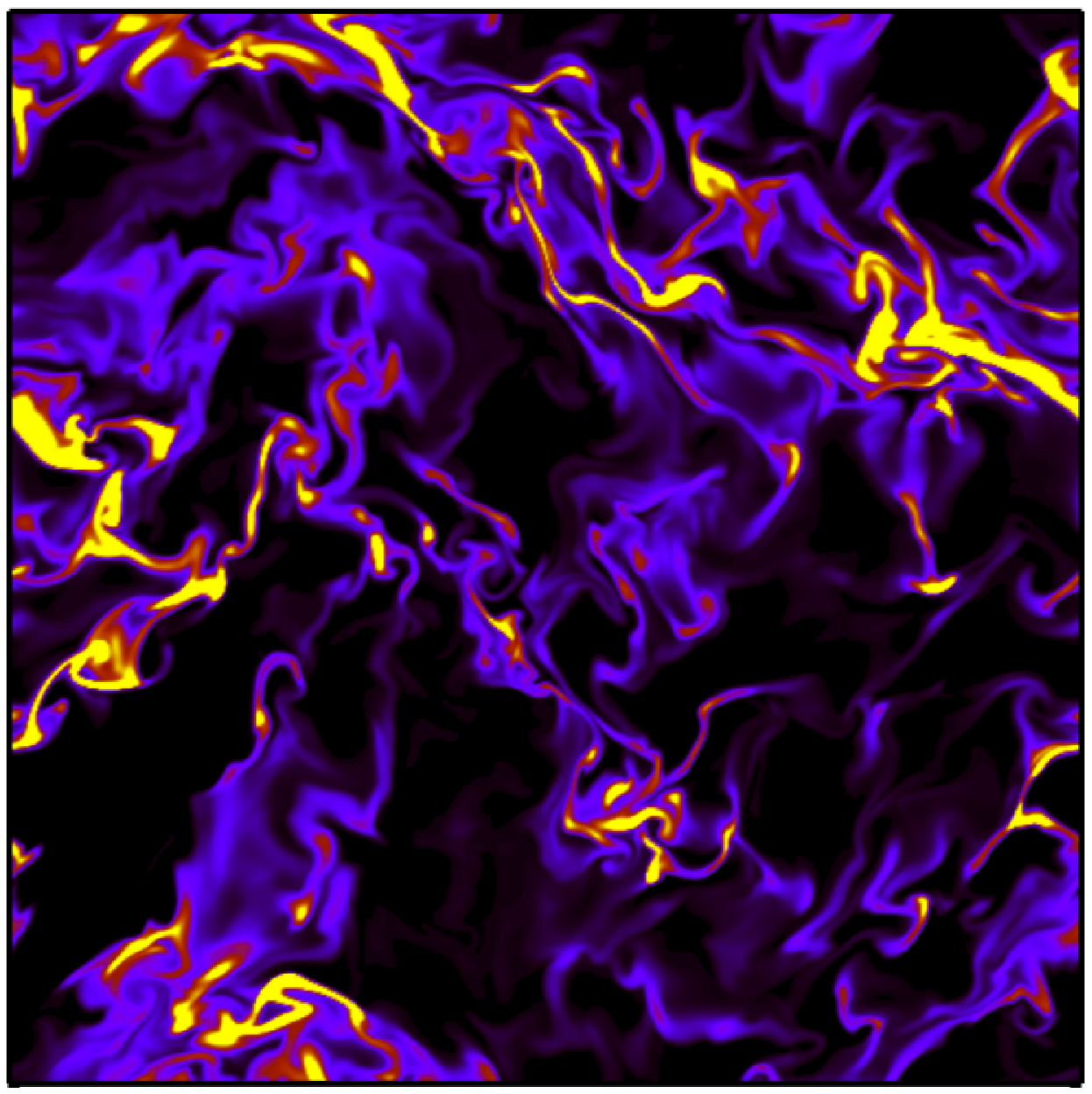}
    \includegraphics[width=0.45\linewidth]{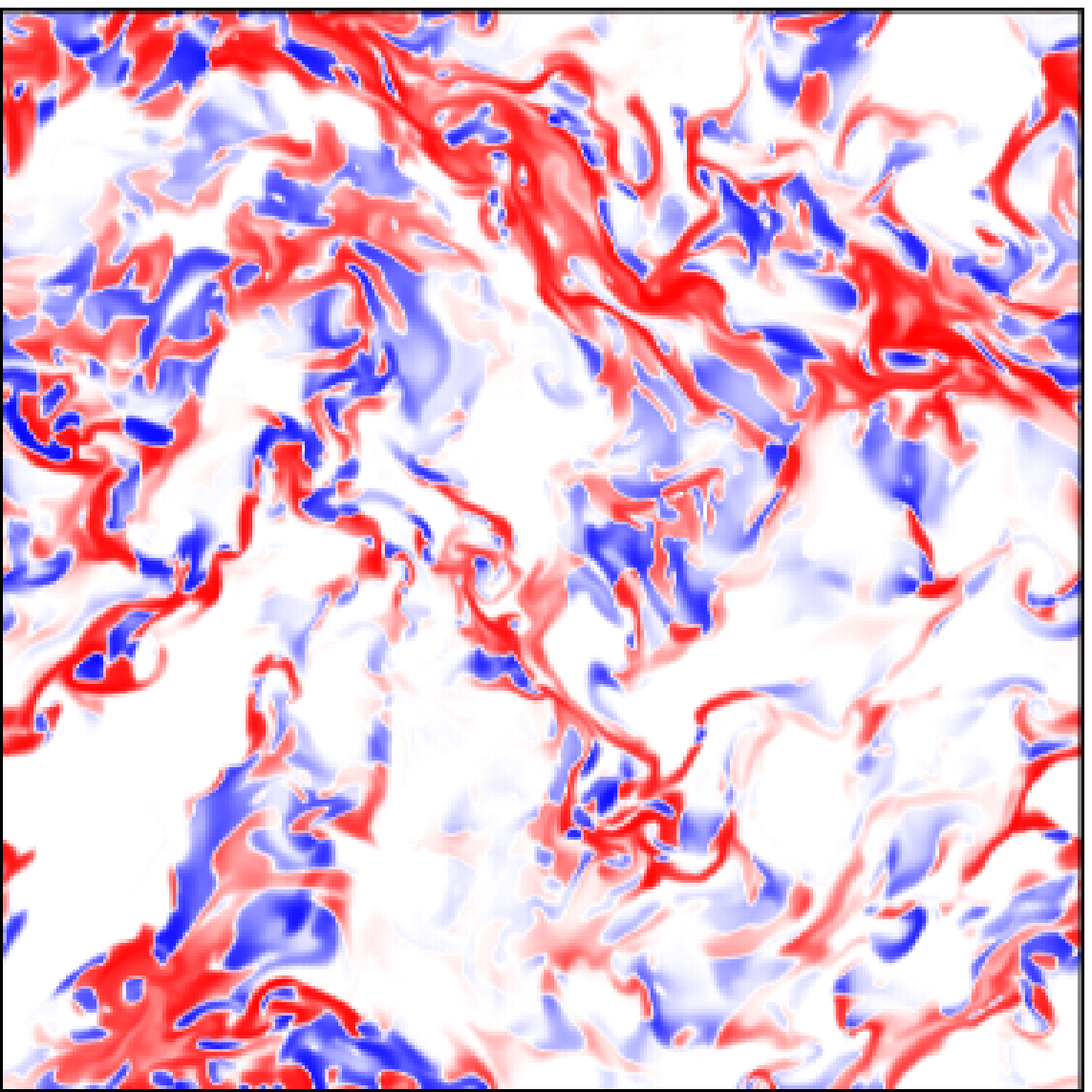}
    \end{center}
    \caption{(Color online) {(Left panel) Pseudocolor plot of
        concentration field. The bright yellow regions indicate
        regions of high concentration ($C>0.1$) and the black regions
        indicate regions of low concentration.  (Right panel)
        Pseudocolor plot of $[C({\bf x},t_0)/(0.1+C({\bf x},t_0))
        (\tanh(-\nabla \cdot {\bf u}))]$.  The dark red regions
        indicate negative divergence and large concentration whereas
        dark blue regions indicate positive divergence and large
        concentration.  Plot are made at identical time $t_0$ (after
        the steady state has been reached) on a slice $z=const$
        obtained from our $512^2$ numerical simulations of
        Eq. (\ref{eq:fish}) for $\mu\tau_{\eta}=0.0045$ and Schmidt
        number $Sc=5.12$. Note that microorganisms cluster near
        regions of compression ($\nabla\cdot {\bf u} < 0$), as is
        evident from the high density of red regions.}}
  \label{fig:fig1}
\end{figure*}

As discussed in \cite{1d}, an advecting compressible turbulent flow
leads to highly non-trivial dynamics. Although the results of
\cite{1d} were obtained only in one dimension using a synthetic
advecting flow from a shell model of turbulence, two striking effects
were observed: the concentration field $C({\bf x},t)$ is strongly
localized near transient but long-lived sinks of the turbulent flows
for small enough growth rate $\mu$; in the same limit, the space-time
average concentration (denoted in the following as carrying capacity)
becomes much smaller than its maximum value $1$.  Both effects are
relevant in biological applications \cite{mur93}.

In this Letter, we present new numerical results for more realistic
two dimensional turbulent flows.  We assume that the microorganism
concentration field $C({\bf x},t)$, whose dynamics is described by
Eq. (\ref{eq:fish}), lies on a planar surface of constant height in a
three dimensional fully developed turbulent flow with periodic
boundary conditions. Such a system could be a rough approximation to
photosynthetic microorganisms that actively control their bouyancy to
manitain a fixed depth below the surface of a turbulent fluid
\cite{mar03}.  As a consequence of this choice, the flow field in the
two dimensional slice becomes compressible \cite{boffetta}.  We
consider here a turbulent advecting field ${\bf u}({\bf x},t)$
described by the Navier-Stokes equations, and nondimensionalize time
by the Kolmogorov dissipation time-scale
$\tau_{\eta}\equiv(\nu/\epsilon)^{1/2}$ and space by the Kolmogorov
length-scale $\eta\equiv(\nu^3/\epsilon)^{1/4}$, where $\epsilon$
  is the mean rate of energy dissipation and $\nu$ is the kinematic
  viscosity. The non-dimensional numbers charecterizing the evolution
of the scalar field $C({\bf x},t)$ are then the Schmidt number
$Sc=\nu/D$ and the non-dimensional time $\mu\tau_{\eta}$.  A
particularly interesting regime arises when the doubling time 
  $\tau_g \equiv \mu^{-1}$ is somewhere in the middle of the inertial
  range of eddy turnover times ($\tau_{r}=r/\delta_r u$, where
$\delta_r u$ is the typical velocity difference across length scale
$r$) that characterize the turbulence. Although the underlying
turbulent energy cascade is somewhat different \cite{mck09}, this
situation arises for oceanic cyanobacteria and phytoplankton, who
double in $8-12$ hours, in a medium with eddy turnover times varying
from minutes to months \cite{mar03}.

The main results of our investigation are the following: we confirm
the qualitative behaviour found in \cite{1d} for a two-dimensional
population, under more realistic turbulent flow conditions. We also
investigate the limit $\mu \to \infty$ and discuss the singular limit
of $\mu \to 0$ validating the physical picture proposed in
\cite{1d}. Our understanding of the limit $\mu \tau_{\eta} \ll 1$ may
be helpful in future investigations, as explicit computations for $\mu
> 0$ can be very demanding. In addition, we quantify the statistical
properties of the concentration field and investigate the effect of an
uniform convective background flow field.

We conducted a three dimensional direct numerical simulation (DNS) of
homogeneous, isotropic turbulence at two different resolutions
($128^3$ and $512^3$ collocation points) in a cubic box of length
$L=2\pi$.  The Taylor microscale Reynolds number \cite{frisch} for the
full 3D simulation was $Re_{\lambda}=75$ and $180$, respectively, the
dimensionless viscosities were $\nu=0.01$ and $\nu=2.05 \cdot
10^{-3}$, the total energy dissipation rate was around $\epsilon
\simeq 1$ in both cases. For the analysis of the Fisher equation we
focused only on the time evolution of a particular 2D slab taken out
of the full three dimensional velocity field and evolved a
concentration field $C({\bf x},t)$ constrained to lie on this plane
only. This is a particularly efficient way of producing a
  compressible 2d velocity field in order to mimic the flow at the
  surface of oceans. Note that our velocity field
  has nothing to do with 2d turbulence and has the structures,
  correlations and spectra of a bidimensional cut of a fully 3d
  turbulent flow.  A typical plot of the $2d$ concentration field and
the concentration field conditioned by the corresponding velocity
divergence field (taken at time $t=86$, $Re_{\lambda}=180$) in this
plane is shown in Fig.~\ref{fig:fig1} ($Sc=5.12$).  The Fisher
equation was stepped forward using a second-order Adams-Bashforth
scheme. The spatial derivatives in the diffusion operator are
discretized using a central, second-order, finite-difference method.
As the underlying flow field is compressible, sharp gradients in the
concentration field can form during time-evolution. In order to
capture these sharp fronts we use a Kurganov-Tadmor scheme for the
advection of the scalar field by the velocity field~\cite{kur00}.

The concentration $C({\bf x},t)$ is highly peaked in small areas,
resembling one dimensional filaments (see Figure \ref{fig:fig1} and
supplemental movie). When the microorganisms grow faster than the
turnover times of a significant fraction of the turbulent eddies,
$C({\bf x},t)$ grows in a quasi-static compressible velocity field and
accumulates near the regions of compression, leading to filaments
\footnote{In two-dimensions, locally, the flow can be characterized by
  determinant and the divergence of the velocity gradient tensor
  $\nabla {\bf u}$.  Negative (positive) values of $\nabla \cdot {\bf
    u}$ correspond to regions of compression (expansion) whereas
  negative (positive) values of the $det(\nabla  {\bf u})$
  correspond to regions of strain (vorticity).}.  The geometry of the
concentration field suggests that $C({\bf x},t)$ is different from
zero on a set of fractal dimension $d_F$ much smaller than $2$. A box
counting analysis of the fractal dimension of $C({\bf x},t)$ supports
this view and provides evidence that $d_F = 1.0 \pm 0.15$.

A biologically important quantity is the spatially averaged carrying
capacity or the density of biological mass in the system,
\begin{equation}
  Z(t) = \frac{1}{L^2}\int dx dy C({\bf x};t), 
  \label{eq:carc}
\end{equation}
and in particular its time average in the statistical steady state as
a function of the growth rate $\mu$, $\langle Z\rangle_{\mu}$.
Without turbulence $\langle Z\rangle_{\mu} =1$ for any $\mu$. When
turbulence is acting, in the limit $\mu \rightarrow \infty$ we expect
the carrying capacity attains its maximum value $\langle Z
\rangle_{\mu \rightarrow \infty} = 1$, because when the characteristic
time $\tau_g $ becomes much smaller than the Kolmogorov dissipation
time $\tau_{\eta}$, the velocity field is a relatively small
perturbation on the rapid growth of the microorganisms.  Indeed,
consider a perturbation expansion of $C({\bf x},t)$ in terms of
$\tau_g$. We define $C({\bf x},t) \equiv \displaystyle
\sum_{n=0,\dots,\infty} \tau_g^n C_n({\bf x},t)$, and substitute in
Eq.~(\ref{eq:fish}), where the functions $C_n({\bf x},t)$ are the
coefficients of the expansion. Upon assuming a steady state and collecting
the terms up to ${\cal O}(\tau_g^2)$ we find, after some algebra,

$\langle Z\rangle_{\mu} \approx 1 - (\tau_g^2/L^2) \langle \int ({\bm
  \nabla} \cdot {\bf u})^2 {\rm {d \bf x}}\rangle + {\cal
  O}(\tau_g^3)$.

The limit $\mu \rightarrow 0$ can be investigated by noting that for
$\mu=0$, Eq. (\ref{eq:fish}) reduces to the Fokker-Planck equation
describing the probability distribution ${\C P({\bf x},t)}$ to find a
Lagrangian particle subject to a time varying force field ${\bf
  u}({\bf x},t)$:
\begin{equation}
  \label{fokkerplanck}
  {\frac{\partial {\C P}}{\partial t}} +  {\bm \nabla} \cdot ({\bf u} {\C P}) = D \nabla^2 {\C P}.
\end{equation}

\begin{figure*}[!t]
  \begin{center}
    \includegraphics[width=0.46\linewidth]{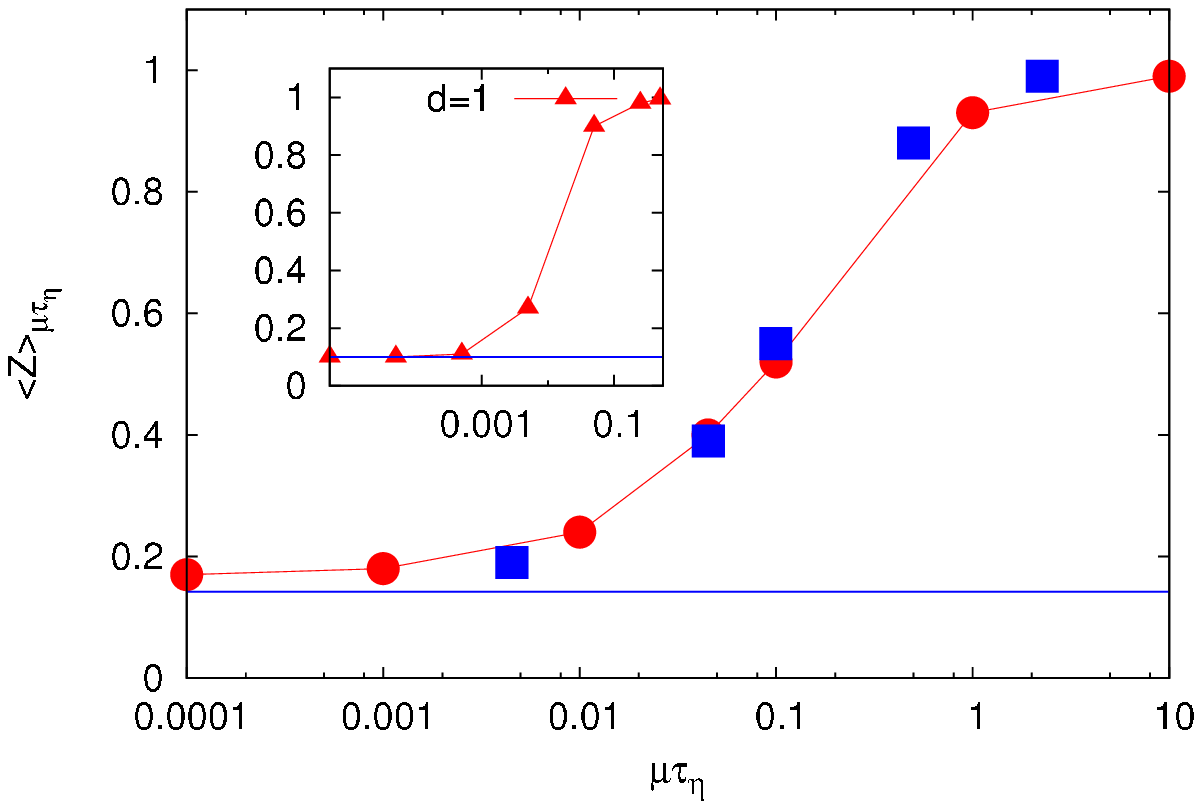}
    \includegraphics[width=0.46\linewidth]{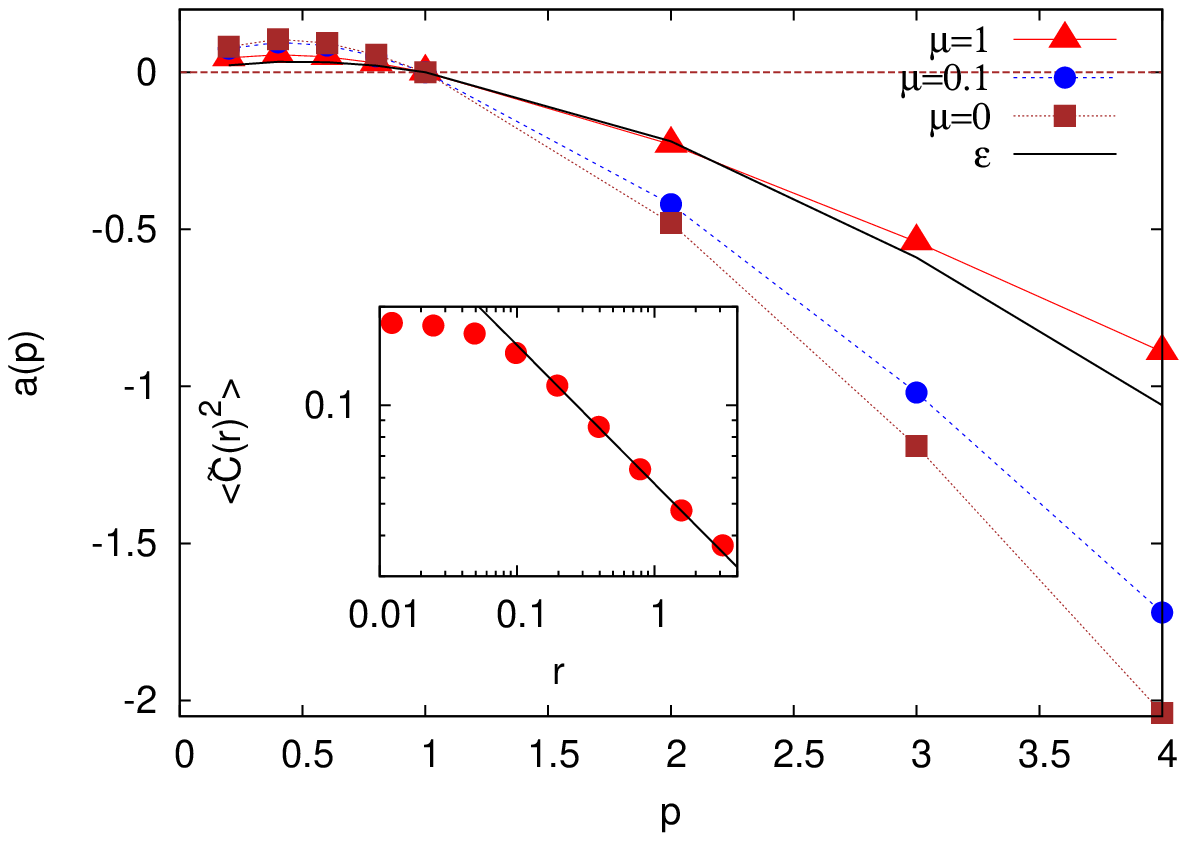}
  \end{center}
  \caption{(Color online) Left panel: Behavior of the carrying
    capacity $\langle Z \rangle_{\mu \tau_{\eta}} $ as a function of
    $\mu \tau_{\eta}$ from $128^2$ (red dots) and $512^2$ (blue
    squares) numerical simulations with $Sc=1$.  Note that for
    $\mu\tau_{\eta} \lesssim 0.001$, the carrying capacity approaches
    the limit $1/(\langle {\C P}^2 \rangle L^4) \approx 0.16\pm 0.02 $
    (blue line) predicted by Eq. (\ref{ZP2}). In the inset we show
    similar results for one dimensional compressible turbulent flows
    in \cite{1d}. Right panel: The anomalous exponents $a(p)$ computed
    by the multifractal analysis of the concentration field $C({\bf
      x},t)$ for different values of $\mu$ and $Sc=1$ for our $512^2$
    numerical simulation.  Note that for $\mu \rightarrow 0$ the
    multifractal exponents approach the statistical properties of the
    field ${\C P}$ described by Eq. (\ref{fokkerplanck}). The black
    line shows the multifractal properties of the energy dissipation
    rate $\epsilon$.  In the inset we show the scaling behavior of
    $\langle \tilde{C}(r)^2 \rangle$ for $\mu=0.1$.}
  \label{figura_mu}
  \label{multifract}
\end{figure*}

Upon defining $\Gamma \equiv \langle ({\bm \nabla} \cdot {\bf u})^2
\rangle^{1/2}$ as the r.m.s value of the velocity divergence,
following \cite{1d} we expect a crossover in the behavior of $\langle
Z \rangle_{\mu}$ for $\mu < \Gamma$. In the limit $\mu \rightarrow 0$,
we expect:
\begin{equation}
  \label{ZP2}
  \displaystyle \lim_{\mu \rightarrow 0 } \langle Z \rangle_{\mu} = \frac{1}{\langle {\C P}^2 \rangle L^4}.
\end{equation}
To understand Eq. (\ref{ZP2}), note first that for small $\mu$ the
statistical properties of $C$ should be close to those of ${\C
  P}$. Thus we can assume that, in a statistical sense, $C({\bf x},t)
\approx L^2 \langle Z \rangle_{\mu} {\C P}({\bf x},t)$. Averaging
Eq. (\ref{eq:fish}) in space and time leads to $\langle C
\rangle_{\mu} - \langle C^2 \rangle_{\mu} = 0$, which is equivalent to
Eq. (\ref{ZP2}). Eq. (\ref{ZP2}) is crucial, because it allows us to
predict $\langle Z\rangle_{\mu}$ for small $\mu$ from the knowledge of
the well-studied statistical properties of Lagrangian tracer particles without
growth in compressible turbulent flows. We have therefore tested both
Eq.~(\ref{ZP2}) and the limit $\mu\rightarrow \infty$ against our
numerical simulations.  In Fig. \ref{figura_mu} we show the behavior
of $\langle Z \rangle_{\mu}$ for the numerical simulations discussed
in this Letter. The horizontal line represents the value $ 1/(\langle
{\C P}^2 \rangle L^4)$ obtained by solving Eq. (\ref{fokkerplanck})
for the same velocity field and $\mu = 0$. The insert shows a similar
result for a one dimensional compressible flow \cite{1d}.  For our
numerical simulations we observe, for $\mu \tau_{\eta}>\Gamma
\tau_{\eta}\approx 0.23$ the carrying capacity $ \langle
Z\rangle_{\mu}$ becomes close to its maximum value $1$.  The limit
$\mu \rightarrow 0$ requires some care: the effect of turbulence is
relevant for $\tau_g$ longer than the Kolmogorov dissipation time
scale $\tau_{\eta}$.  We take the limit $\mu\to 0$ at {\it{fixed}}
system size $L$.  When $\tau_g \gg \tau_L\sim (L^2/\epsilon)^{1/3}$, 
the large scale correlation time, the population is effectively frozen
on all turbulent time scales, and Eq. (\ref{ZP2}) should apply.

The limit $\mu \rightarrow 0$ can be investigated more precisely as
follows: according to known results on Lagrangian particles in
compressible turbulent flows, ${\C P}$ should have a multifractal
structure in the inviscid limit $\nu \rightarrow 0$ \cite{bec03,
  massimo, boffetta}.  If our assumption leading to Eq. (\ref{ZP2}) is
correct, $C({\bf x},t)$ must show multifractal behavior in the same
limit with multifractal exponents similar to those of ${\C P}$.

We perform a multifractal analysis of the concentration field
$C(x,y,t)$ with $\mu>0$ by considering the average quantity
$\tilde{C}_{\mu}(r,t) \equiv \frac{1}{r^2} \int_{B(r)} C(x,y,t) dx dy
$ where $B(r)$ is a square box of size $r$. Then the quantities
$\langle \tilde{C}_{\mu}(r)^p \rangle $ are expected to be scaling
functions of r, i.e.  $ \langle \tilde{C}_{\mu}(r)^p \rangle \sim
r^{a(p)}$, where $a(p)$ is a non linear function of $p$.

In Figure \ref{multifract} we show the quantity $a(p)$ for $\mu=0,
0.1,$ and $1$ for $0\le p \le 4$ extracted from power law fits to
$\langle \tilde{C}(r)^p \rangle$ over $\sim 1.5$ decades. In the inset
we show $\langle \tilde{C}(r)^2 \rangle $ for $\mu=0.01$. Although our
dynamic range is limited, the scaling description seems to work with
smoothly varying exponents $a(p)$. Even more important, the
statistical properties of $\tilde{C}_{\mu}(r)$ seems to converge to
the case $\mu=0$ as $\mu\to 0$.  In the same figure we also show, for
comparison, a similar analysis performed for the energy dissipation
field (black line); see \cite{frisch} for a detailed description.

\begin{figure}[!b]
\vskip -0.5cm
  \begin{center}
    \includegraphics[width=1.\hsize]{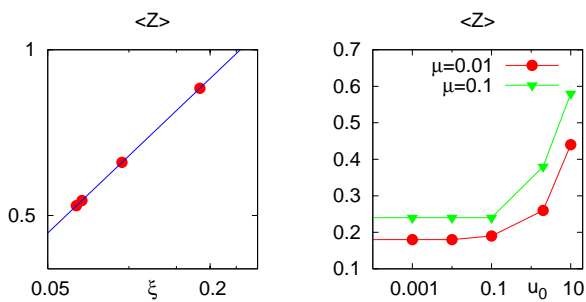}
  \end{center}
  \caption{(Color online) Left: Log-log plot of $\langle Z \rangle$ as
    a function of the localization length $\xi$ defined in
    Eq. (\ref{xi}) for $u_0=0$.  The blue line is the fit to the
    power-law.  The slope is consistent with the prediction $\langle
    Z\rangle \sim \xi^{-a(2)}$ discussed in the text. The numerical
    simulations are done for $\mu=0.01$ and different values of $D$
    from $D=0.05$ to $D=0.001$.  Right: Plot of $ \langle Z \rangle$
    as function of a super-imposed uniform velocity $u_0$ for
    $\mu=0.01$ (red bullets) and $\mu=0.1$ (green triangles) with 
    $D=0.015$.}
  \label{zversusxi}
\end{figure}

Our multifractal analysis suggests a relation between the quasi
localization length $\xi$ and the carrying capacity $\langle Z
\rangle_{\mu}$.  The quasi localization length $\xi$ can be considered
as the smallest scale below which one should not observe fluctuations
of $C({\bf x},t)$. In the limit $\mu \to 0$, we can define the quasi
localization length $\xi$ as:
\begin{equation}
  \label{xi}
  \xi^2 \equiv   \frac{ \langle { \C P }^2 \rangle} {  \langle  (\nabla {\C P}) ^2 \rangle}.
\end{equation}

We expect $\xi$ to be of the same order of the width of the narrow
filaments in Fig. \ref{fig:fig1}. To compute $\langle Z\rangle_{\mu}$
as a function of $\xi$, we observe that it is reasonable to assume
$\langle {\C P}^2 ({\bf x},t) \rangle \sim \langle C^2(r=\xi,t)
\rangle \sim \xi^{a(2)}$.  Using Eq. (\ref{ZP2}) we obtain $\langle Z
\rangle_{\mu} \sim \xi^{-a(2)}$.  On the left side of Figure
\ref{zversusxi} we show $\langle Z \rangle_{\mu}$ as a function of
$\xi$ (obtained by using (\ref{xi})) for $\mu=0.01$ by varying the
diffusivitiy $D$. Reducing the diffusivity $D$ shrinks the
localization length $\xi$ and $\langle Z \rangle_{\mu}$ becomes
smaller.  Dimensional analysis applied to 
Eq. \ref{fokkerplanck} and \ref{xi} suggests that 
$\xi \propto (D^2\nu/\epsilon)^{1/4}$. More generally we expect that $\xi(D)$ is
  monotonically increasing with $D$. From Figure \ref{zversusxi} (left side), 
a reasonable power law behavior is observed with a scaling exponent 
$0.46 \pm 0.03$ very close to the predicted behavior $-a(2) = 0.47$ 
obtained from Fig. \ref{multifract} (right side inset).

Note that Fig. \ref{zversusxi} (left side) reveals a strong dependence
of $\langle Z \rangle_\mu$ on $\xi$ and hence on the microbial
diffusion constant. $D$ in turn depends on the ability of marine
microorganisms to swim. Approximately $1/3$ of the open ocean isolates
of {\it Synechococcus} can propel themselves along their micron-sized
long axis at velocities of $\sim 25 \mu m/sec$ \cite{ehl96}.  Upon
assuming a random direction change every $\sim 20-30$ body lengths,
the effective diffusion constant entering Eq.~(\ref{eq:fish}) can be
enhanced $1000-$fold relative to $D$ for passive organisms. The
extensive energy investment required for swimming in a turbulent ocean
becomes more understandable in light of the increased carrying
capacity associated with say, a $\sim 30-$fold increase in $\xi$. Some
marine microorganisms may have evolved swimming in order to mitigate
the overcrowding associated with compressible turbulent advection.

Finally, we discuss bacterial populations subject to both turbulence
and uniform drift because of, e.g., the ability to swim in a
particular direction. In this case, we can decompose the velocity
field into zero mean turbulence fluctuations plus a constant drift
velocity $u_0$ \cite{sp07} along e.g. the $x$-direction. In presence
of a mean drift velocity Eq.~(\ref{eq:fish}) becomes:
\begin{equation}
  {\frac{\partial C}{\partial t}} + {\bm \nabla} \cdot [({\bf u} + u_0 \hat{\bm e_x})C] = D \nabla^2 C +  \mu C(1 -  C)
\end{equation}
where $\hat{\bm e_x}$ is the unit vector along the $x$-direction. Note
that a mean drift (to follow nutrient gradients, say) breaks the
Galilean invariance as the concentration $C$ is advected by $u_0$,
while turbulent fluctuations ${\bf u}$ remain fixed. In
Fig.~\ref{zversusxi} we show the variation of carrying capacity versus
$u_0$ for two different values of $\mu$ and fixed diffusivity
$D=0.015$. We find that for $u_0 \lesssim u_{rms}$ (the
root-mean-square turbulent velocity) the carrying capacity $Z$
saturates to a value equal to the value of $Z$ in absence of $u_0$
i.e., quasilocalization by compressible turbulence dominates the
dynamics. For $u_0>u_{rms}$ the drift velocity delocalizes the
bacterial density eventually causing $Z \to 1$, as was also found in
$d=1$.

We have shown that a realistic model for two dimensional compressible
turbulence predicts reduced microorganism carrying capacities, similar
to those found in a highly simplified $1d$ model
\cite{1d}. Simulations at two elevated Reynolds number show that
results are robust and in agreement when properly normalized. The
limit of large growth rates was addressed analytically, and the
statistics maps smoothly onto known results for conserved densities
advected by compressible turbulence.  Finally we studied the effect of
a preferred swimming direction on the carrying capacity.

{\bf Acknowledgment}
We thank M.H. Jensen, A. Mahadevan, S. Pigolotti, and A. Samuel for
useful discussions.  We acknowledge computational support from CASPUR
(Roma, Italy uner HPC Grant 2009 N. 310), from CINECA (Bologna, Italy)
and SARA (Amsterdam, The Netherlands). Support for D.R.N. was provided
in part by the National Science Foundation through Grant DMR-0654191
and by the Harvard Materials Research Science and Engineering Center
through NSF Grant DMR-0820484.  Data from this study are publicly
available in unprocessed raw format from the iCFDdatabase
(http://mp0806.cineca.it/icfd.php).

\end{document}